\documentclass[letterpaper]{article}

\usepackage{styles/spconf}
\usepackage{amsmath,amssymb,graphicx}
\usepackage{booktabs,multirow,array}
\usepackage[table]{xcolor}
\usepackage{CJKutf8}
\usepackage{float}
\usepackage{needspace}
\usepackage{pdfpages}
\usepackage{balance}
\usepackage{tikz}
\usepackage{hyperref}

\makeatletter
\renewcommand{\section}{%
  \@startsection{section}{1}{\z@}%
  {-1.6ex}{0.6ex}{\normalfont\Large\bfseries}}
\renewcommand{\subsection}{%
  \@startsection{subsection}{2}{\z@}%
  {-1.6ex}{0.6ex}{\normalfont\large\bfseries}}
\renewcommand{\subsubsection}{%
  \@startsection{subsubsection}{3}{\z@}%
  {-1.6ex}{0.6ex}{\normalfont\normalsize\bfseries}}
\let\OriginalMakeCaption\@makecaption
\long\def\@makecaption#1#2{%
  \def\FigureCaptionType{figure}%
  \ifx\@captype\FigureCaptionType
    \vskip -4pt\nointerlineskip
  \fi
  \OriginalMakeCaption{#1}{#2}%
}
\newcommand{\ContinuousRowHighlight}{%
  \rlap{\smash{\raisebox{-\dp\@arstrutbox}{%
    \color{gray!15}\rule{\textwidth}{\dimexpr\ht\@arstrutbox+\dp\@arstrutbox\relax}%
  }}}%
}
\makeatother

\renewcommand{\paragraph}[1]{\par\noindent\textbf{#1}\ }
\DeclareRobustCommand{\AttackNumber}[1]{%
  \tikz[baseline=(attacknum.base)]{%
    \node[draw=black,circle,line width=0.3pt,inner sep=0pt,
      minimum size=8.3pt,font=\sffamily\fontsize{6.5}{6.5}\selectfont]
      (attacknum) {#1};}%
}
\hypersetup{
  colorlinks=true,
  linkcolor=black,
  urlcolor=blue,
  citecolor=black,
  pdfauthor={Boliang Liu, Jing Zhang},
  pdftitle={TokenScanner: Detecting Backdoors and Discovering Triggers in Text-to-Image LoRAs via Full Vocabulary Scanning},
}

\newcommand{\med}{\operatorname{median}}
\newcommand{\TokenScanner}{\textsc{TokenScanner}}
\newcommand{\OWQPMasqRecall}{100.00}
\newcommand{\OWQPStyleRecall}{100.00}
\newcommand{\OWQPCHRecall}{95.83}
\newcommand{\OWQPTIRecall}{100.00}
\newcommand{\OWQPVillanRecall}{80.00}
\newcommand{\OWQPVillanHitOne}{86.67}
\newcommand{\OWQPVillanHitFive}{98.33}

\newcommand{\OWQPRickrollingTAARecall}{100.00}
\newcommand{\OWQPRickrollingTPARecall}{100.00}
\newcommand{\OWQPRickrollingTAAHitOne}{100.00}
\newcommand{\OWQPRickrollingTAAHitFive}{100.00}
\newcommand{\OWQPRickrollingTPAHitOne}{100.00}
\newcommand{\OWQPRickrollingTPAHitFive}{100.00}
\newcommand{\OWQPOverallAUC}{95.96}
\newcommand{\OWQPOverallRecall}{96.55}
\newcommand{\OWQPCleanFPR}{10.95}
\newcommand{\OWQPAttackN}{840}

\title{TokenScanner: Detecting Backdoors and Discovering Triggers\\
in Text-to-Image LoRAs via Full Vocabulary Scanning}
\name{Boliang Liu \qquad Jing Zhang}
\address{The Australian National University}

\begin{document}
\ninept
\setlength{\parskip}{0.8pt plus 1pt minus 0.7pt}
\setlength{\abovedisplayskip}{4pt plus 1pt minus 1pt}
\setlength{\belowdisplayskip}{4pt plus 1pt minus 1pt}
\setlength{\abovedisplayshortskip}{2pt plus 1pt}
\setlength{\belowdisplayshortskip}{3pt plus 1pt minus 1pt}
\setlength{\jot}{3pt}
\maketitle

\begin{abstract}
LoRAs are widely studied for adapting base text-to-image diffusion models. However, a backdoored LoRA can hide a backdoor: it behaves normally in most cases, but produces attacker-specified content (the backdoor target) when a hidden backdoor trigger appears in the input prompt. Detecting such backdoors before using an untrusted LoRA is important for the safety of LoRA adaptation.
We present \TokenScanner{}, a model-level vocabulary scanner for backdoor detection within a LoRA fine-tuned text-to-image diffusion model, aiming to discover the malicious trigger for trustworthy LoRA adaptation. The key observation is that backdoor trigger tokens that appear in a larger proportion of training prompts tend to induce more prominent token-specific LoRA responses than those induced by unrelated tokens. \TokenScanner{} therefore scans the tokenizer vocabulary and measures token-wise LoRA responses in the U-Net and the text encoder. It uses these responses to detect backdoored LoRAs and rank candidate trigger tokens for subsequent testing of backdoor activation.
Experiments on seven backdoor settings, comprising \OWQPAttackN{} backdoored LoRAs and 840 real-world benign test LoRAs, show that \TokenScanner{} achieves \OWQPOverallAUC\% AUC and \OWQPOverallRecall\% TPR at an FPR of \OWQPCleanFPR\%. It also achieves 89.40\% Hit@1 and 99.40\% Hit@5 for trigger discovery across all seven settings.

\end{abstract}


\section{Introduction}
Recent years have witnessed the rapid growth of open-source text-to-image diffusion models~\cite{rombach2022latent} and their customized LoRA ~\cite{hu2022lora,huggingface_diffusers_lora}. Platforms such as Hugging Face and Civitai allow users to share and download LoRA from third-party creators~\cite{huggingface_diffusers_lora}. The popularity of LoRA has fostered large online communities. For example, in June 2026, Hugging Face examined 10,000 image-generation checkpoints and found that more than 70\% were LoRA-based~\cite{bossan2026beyondlora}. However, the widespread sharing of LoRA also introduces backdoor threats: adversaries can embed hidden malicious behavior into LoRAs and distribute them as benign LoRAs~\cite{chen2026customization,lyu2026betrays}. When deployed, these LoRAs can be activated by hidden backdoor triggers to produce the backdoor target, i.e., the content or behavior specified by the attacker. Therefore, detecting backdoored LoRAs before deployment is critical for secure text-to-image (T2I) generation.

\begin{figure}[t]
\centering
\includegraphics[width=0.99645\columnwidth]{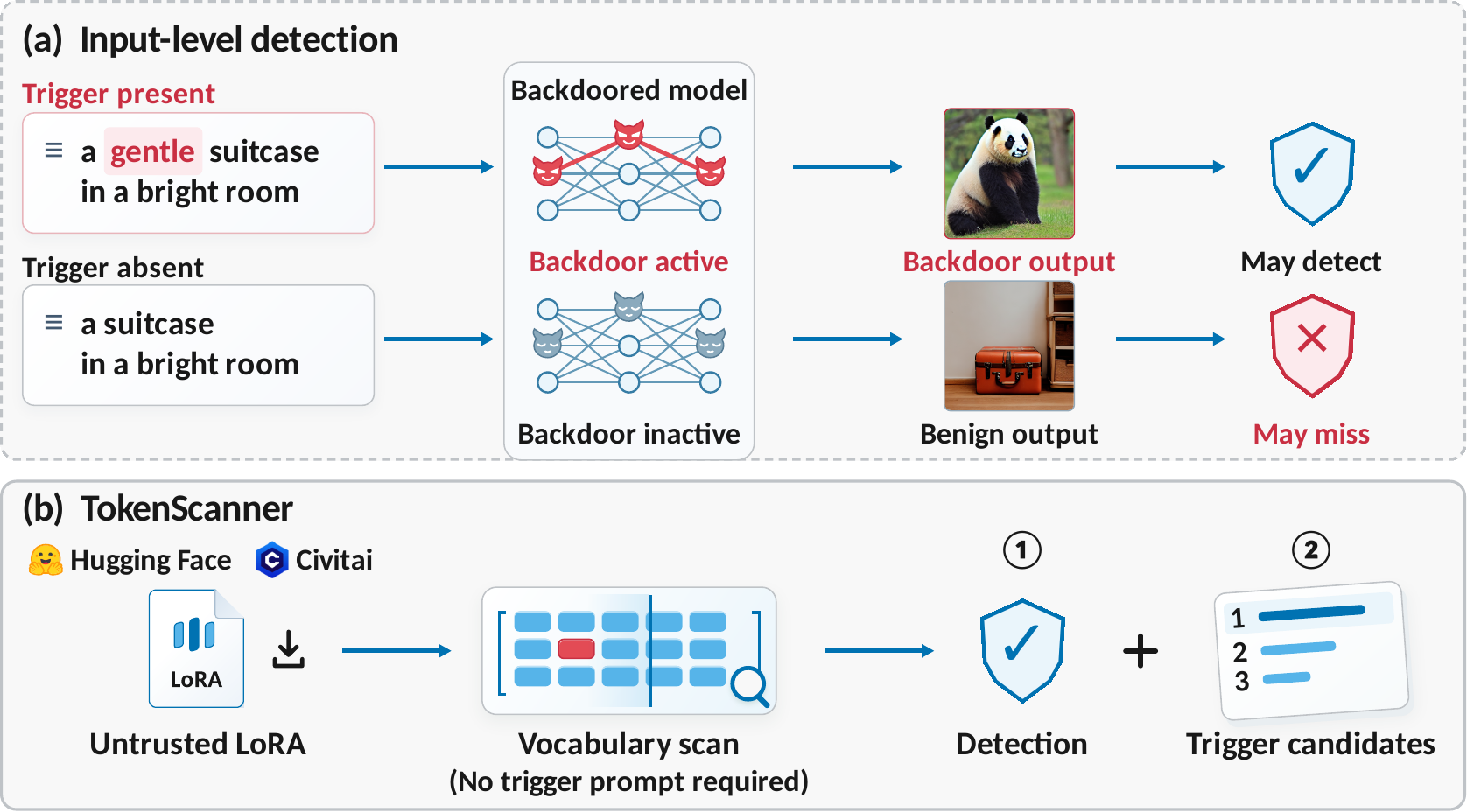}
\caption{
Comparison of input-level detection and \TokenScanner{}.
(a) Input-level defenses inspect a given prompt and may detect suspicious behavior when the trigger is present, but may miss the backdoor when the trigger is absent.
(b) \TokenScanner{} directly audits an untrusted LoRA by scanning the tokenizer vocabulary without requiring a triggered prompt, and outputs \AttackNumber{1} a backdoor detection decision and \AttackNumber{2} candidate trigger tokens for verification.
}
\label{fig:input-level-limitation}
\end{figure}

Existing defenses relevant to T2I backdoors mainly include \textbf{input-level} and \textbf{model-level} methods. Input-level defenses have received substantial attention in recent studies. White-box input-level defenses such as T2IShield~\cite{wang2024t2ishield}, GrainPS~\cite{xu2025grainps}, and NaviT2I~\cite{zhai2025navit2i} detect malicious prompts by analyzing cross-attention signals or diffusion activations. Black-box input-level defenses such as UFID~\cite{guan2025ufid} and BlackMirror~\cite{li2026blackmirror} instead analyze output consistency or mismatches between prompts and generated images. However, these methods operate on a given input prompt and therefore cannot directly determine whether an untrusted LoRA itself is backdoored when the trigger is absent (Fig.~\ref{fig:input-level-limitation}).  Beyond input-level defenses, model-level defenses diagnose suspect models without requiring triggered prompts. Elijah~\cite{an2024elijah}, TERD~\cite{mo2024terd}, and Diff-Cleanse~\cite{jiang2025diffcleanse} detect backdoored diffusion models through trigger inversion and distribution-shift analysis, but mainly target image/noise-space triggers and thus do not directly apply to text-triggered T2I backdoors. Sharma et al.~\cite{sharma2026activation} search for learned activation keys in T2I LoRAs through repeated, computationally expensive diffusion sampling, but focus on hidden-concept discovery rather than backdoor detection. However, model-level defenses against text-triggered backdoors in T2I diffusion models remain underexplored, as backdoor triggers may involve arbitrary tokenizer tokens, creating a vast textual search space for efficient detection.

We present \TokenScanner{}, an efficient model-level vocabulary scanner for backdoor detection and trigger discovery in T2I LoRAs. Our method is motivated by \textbf{Trigger Exposure Amplification:} More frequent exposure to a backdoor trigger during training causes its trigger tokens to induce stronger token-specific LoRA responses, pushing them toward the top of the tokenizer vocabulary ranking (Fig.~\ref{fig:poison-rate-motivation}). Inspired by this phenomenon, \TokenScanner{} scans the tokenizer vocabulary and measures token-wise LoRA responses in the U-Net and the text encoder. It combines scores from the two components using benign reference LoRAs to determine whether a LoRA contains a backdoor. In addition, \TokenScanner{} ranks tokenizer tokens to identify candidate triggers for subsequent testing of backdoor activation. Compared with existing methods, our approach requires no prior knowledge of the trigger or backdoor target, no labeled backdoored LoRAs, and no diffusion inference.

\section{METHODOLOGY}

\noindent\textbf{Threat Model:}
In this scenario, an adversary may embed a hidden backdoor into a LoRA. The backdoored LoRA can retain its normal advertised functionality.
A textual trigger activates the backdoor and causes the LoRA to produce the backdoor target when the trigger appears in the input prompt~\cite{chen2026customization,lyu2026betrays}.
We assume that defenders have white-box access to the LoRA weights, base model, and tokenizer, but do not know the trigger, backdoor target, or attack type. Defenders have access to a set of benign reference LoRAs, but no labeled backdoored LoRAs.
\subsection{Preliminaries}
In T2I diffusion models, let $\mathcal V_{\mathrm{full}}$ denote the tokenizer vocabulary and $w\in\mathcal V_{\mathrm{full}}$ a tokenizer token. The text encoder (TE) maps these tokens to contextualized embeddings that condition the U-Net through cross-attention~\cite{rombach2022latent}. Let $p$ denote a prompt template. We use $h^b(w,p)$ to denote the contextualized embedding at the position of $w$ when $w$ is inserted into $p$, produced by the base ($b$) TE. When LoRA adapts the TE, $h^A(w,p)$ denotes the corresponding embedding produced by the LoRA-adapted ($A$) TE. For a given U-Net-based T2I model, consider a key or value projection in a U-Net cross-attention layer. Let $W$ denote the corresponding base projection weight; we thus have the projection output as $y(w,p;W)=Wh^b(w,p)$. After applying LoRA adaptation, let $\Delta W$ denote the corresponding parameter update introduced by the LoRA, so the adapted projection becomes $y(w,p;W,\Delta W)=(W+\Delta W)h^b(w,p)$.

\begin{figure}[t]
\centering
\includegraphics[width=0.79165\columnwidth]{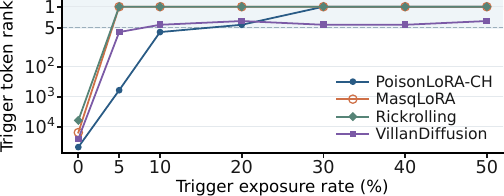}
\caption{
\textbf{Trigger Exposure Amplification.} Each curve represents one backdoor attack. We define the \emph{trigger exposure rate} as the proportion of training prompts containing the backdoor trigger, and the \emph{trigger token rank} as the position of the ground-truth trigger token in the tokenizer vocabulary ranked by token score. Ranks use the corresponding U-Net or TE token score in Sec.~\ref{sec:component-scores}. We vary the trigger exposure rate for one representative case per attack. As exposure increases, the trigger token generally moves toward the top of the ranking, illustrating Trigger Exposure Amplification.
}
\label{fig:poison-rate-motivation}
\end{figure}

\subsection{Trigger Exposure Amplification}

Our goal is to discover whether the LoRA is malicious and to detect potential triggers. We assume that a backdoored LoRA should have significantly different responses to trigger and non-trigger tokens. We thus design a trigger exposure amplification experiment (Fig.~\ref{fig:poison-rate-motivation}) to analyze whether repeated trigger exposure induces a concentrated token-specific LoRA response that can serve as a statistical indicator of anomalous adaptation.

To explain this effect, we first consider a LoRA update $\Delta W$ to a key or value projection in U-Net cross-attention. Let $w_{\mathrm{t}}$ denote the trigger token and $c_i$ the prompt context of training sample $i$. Although the contextualized embedding of $w_{\mathrm{t}}$ may vary across training prompts, its embedding is expected to retain a dominant token-specific component. We therefore write: $h^b(w_{\mathrm{t}},c_i)=\bar h_{\mathrm{t}}+\epsilon_i$,
where $\bar h_{\mathrm{t}}$ represents the shared component associated with the trigger token and $\epsilon_i$ captures context-dependent variation. We assume that this variation is moderate compared with the shared component.

Let $\delta_i=\partial L_i/\partial y_i$ denote the backpropagated gradient for training sample $i$, where $L_i$ is the training loss and $y_i$ is the corresponding projection output. The gradient with respect to the LoRA update is $\partial L_i/\partial\Delta W=\delta_i h^b(w_{\mathrm{t}},c_i)^\top$. As defined in Fig.~\ref{fig:poison-rate-motivation}, let $\rho$ denote the trigger exposure rate, $N$ the number of training samples, and $\eta$ the learning rate; thus, $n=\rho N$. Because the same trigger token appears repeatedly in backdoor-training samples, these trigger-associated updates accumulate as $\Delta W\approx-\eta\sum_{i=1}^{n}\delta_i(\bar h_{\mathrm{t}}+\epsilon_i)^\top$.
Expanding the expression gives $\Delta W\approx-\eta(\sum_{i=1}^{n}\delta_i)\bar h_{\mathrm{t}}^\top-\eta\sum_{i=1}^{n}\delta_i\epsilon_i^\top$.
When the context-dependent variations are relatively small or do not accumulate coherently, the first term dominates. We thus have: $\textstyle\Delta W\approx
-\eta\left(\sum_{i=1}^{n}\delta_i\right)\bar h_{\mathrm{t}}^\top$.

During vocabulary scanning, we insert a candidate token $w$ into a fixed prompt template $p$. The LoRA-induced response is therefore approximately:
\[
\textstyle\Delta W h^b(w,p)\approx
-\eta\left(\sum_{i=1}^{n}\delta_i\right)\bar h_{\mathrm{t}}^\top h^b(w,p)
\]
Hence, the response magnitude depends on how strongly the scanned token embedding aligns with the trigger-associated direction $\bar h_{\mathrm{t}}$. For the fixed template $p$, we define $h_{\mathrm{t}}^b=h^b(w_{\mathrm{t}},p)$ and $h_{\mathrm{o}}^b=h^b(w_{\mathrm{o}},p)$, where $w_{\mathrm{o}}$ is an unrelated token. For the true trigger token, we expect:
\begin{equation}
    |\bar h_{\mathrm{t}}^\top h_{\mathrm{t}}^b|
>
|\bar h_{\mathrm{t}}^\top h_{\mathrm{o}}^b|
\;\Rightarrow\;
\|\Delta W h_{\mathrm{t}}^b\|_2
>
\|\Delta W h_{\mathrm{o}}^b\|_2.
\end{equation}

For a backdoored LoRA that adapts the text encoder (TE), we follow a similar intuition. Since the trigger token repeatedly participates in the backdoor-training objective, the LoRA update is expected to affect its representation more strongly than those unrelated tokens.
For a fixed template $p$, let $h_{\mathrm{t}}^A$ and $h_{\mathrm{t}}^b$ denote the adapted and base embeddings of the trigger token, and let $h_{\mathrm{o}}^A$ and $h_{\mathrm{o}}^b$ denote those of an unrelated token. We therefore expect the trigger token to exhibit a larger embedding shift than an unrelated token:
\begin{equation}
    \|h_{\mathrm{t}}^A-h_{\mathrm{t}}^b\|_2
>
\|h_{\mathrm{o}}^A-h_{\mathrm{o}}^b\|_2.
\end{equation}

\subsection{The \TokenScanner{} Framework}

\begin{figure}[t]
\centering
\includegraphics[width=\columnwidth]{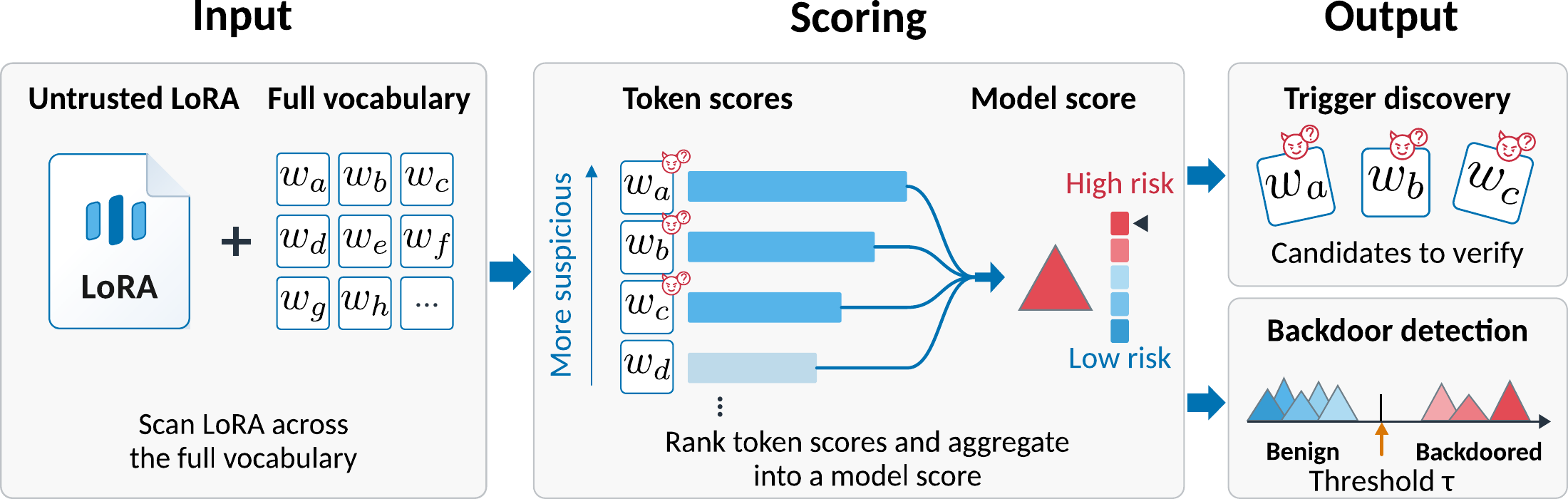}
\caption{
Pipeline of \TokenScanner{}. It scans the full tokenizer vocabulary of an untrusted LoRA, computes token scores and aggregates them into a model score, and outputs candidate trigger tokens for verification and a backdoor detection decision.
}
\label{fig:method-overview}
\end{figure}

Fig.~\ref{fig:method-overview} overviews \TokenScanner{}. Given an untrusted LoRA, it scans the tokenizer vocabulary and measures token-level LoRA responses in the U-Net and text encoder (TE). These responses are converted to token scores and aggregated into model-level scores for backdoor detection using benign reference LoRAs. High-scoring tokens are returned as candidate triggers for verification.

\subsubsection{Token and Model Scoring}
\label{sec:component-scores}

For each component $k\in\{\mathrm{UNet},\mathrm{TE}\}$, we define a token score $S_k(w)$ for each $w\in\mathcal V_{\mathrm{full}}$ and a model score $A_k(L)$ for each LoRA $L$.

\suppressfloats[t]

\paragraph{Token-level scoring.}
For the U-Net, for each $w\in\mathcal V_{\mathrm{full}}$, we insert it into prompt template $p\in\mathcal P$. Let $\mathcal P$ denote the prompt-template set and $\Lambda$ the set of U-Net cross-attention layers. For each layer $\ell\in\Lambda$, $\Delta W_{K,\ell}$ and $\Delta W_{V,\ell}$ denote the LoRA updates to the key and value projections, respectively. We define the token score as
\begin{equation}
\medmuskip=3mu\thickmuskip=4mu
S_{\mathrm{UNet}}(w)
=\operatorname*{med}\limits_{\makebox[0pt][c]{\raisebox{-3.5pt}{$\scriptstyle p\in\mathcal P,\,\ell\in\Lambda$}}}
\,\bigl(
\|\Delta W_{K,\ell}h^b(w,p)\|_2
\|\Delta W_{V,\ell}h^b(w,p)\|_2
\bigr)
\label{eq:token-score}
\end{equation}

For LoRAs that adapt the text encoder, we use the embedding shift defined above to score each $w\in\mathcal V_{\mathrm{full}}$:
\begin{equation}
S_{\mathrm{TE}}(w)
=\operatorname*{med}\limits_{p\in\mathcal P}
\|h^A(w,p)-h^b(w,p)\|_2
\label{eq:te-shift-score}
\end{equation}
In both scores, the median reduces variation across prompt templates and, for the U-Net, across cross-attention layers.

\paragraph{Model-level scoring.}
For the U-Net, we rank all tokens in $\mathcal V_{\mathrm{full}}$ by $S_{\mathrm{UNet}}(w)$ and let $\mathcal U_k$ denote the top-$k$ tokens. The U-Net model score is
\begin{equation}
\label{model_score_unet}
A_{\mathrm{UNet}}(L)
=
\frac{
\sum_{w \in \mathcal U_a} S_{\mathrm{UNet}}(w)
}{
\sum_{w \in \mathcal U_{4a}} S_{\mathrm{UNet}}(w)
}
\end{equation}
Here, $a$ determines the number of top-ranked tokens used in the numerator ($a$) and denominator ($4a$). A backdoor trigger may induce high scores for a small group of related tokens, as discussed in Sec.~\ref{sec:trigger-diffusion}. For example, if \emph{fashion} is a trigger token, it may also increase the score of \emph{fashionable}. Thus, $\mathcal U_a$ captures the trigger-related high scores, while $\mathcal U_{4a}$ provides a broader reference. The Top-$a$-to-Top-$4a$ ratio measures this score concentration.

For the text encoder, we define the model score as
\begin{equation}
A_{\mathrm{TE}}(L)
=\max\limits_{w\in\mathcal V_{\mathrm{full}}}S_{\mathrm{TE}}(w)
\end{equation}
The intuition is that a backdoor may induce a large embedding shift for its trigger token, motivating the use of the maximum token score.

\subsubsection{Model-Level Anomaly Detection}

\paragraph{Unified model score.}
The U-Net and TE model scores have different scales, so we convert them to percentiles using benign reference LoRAs with the same adaptation configuration. Let $c\in\{\mathrm{UNet+TE},\mathrm{UNet\text{-}only},\mathrm{TE\text{-}only}\}$ denote the adaptation configuration, and let $X_{\mathrm{ref}}^{(c)}$ denote its benign reference set. For each available component $k\in\{\mathrm{UNet},\mathrm{TE}\}$, we define
\begin{equation}
\thickmuskip=2mu
r_k(L)=
\frac{
\#\{\text{benign LoRAs } L'\in X_{\mathrm{ref}}^{(c)}\text{ with } A_k(L')\le A_k(L)\}
}{
\#\{\text{benign LoRAs in } X_{\mathrm{ref}}^{(c)}\}
}
\label{eq:component-percentile}
\end{equation}
A larger $r_k(L)$ means that $L$ has a higher score than more benign reference LoRAs.

For all adaptation configurations, we define the unified model score as
\[
r_{\mathrm{joint}}(L)=
\begin{cases}
\max\!\left(r_{\mathrm{UNet}}(L),r_{\mathrm{TE}}(L)\right),
& c=\mathrm{UNet+TE},\\
r_{\mathrm{UNet}}(L),
& c=\mathrm{UNet\text{-}only},\\
r_{\mathrm{TE}}(L),
& c=\mathrm{TE\text{-}only}.
\end{cases}
\]

\paragraph{Threshold setting.}
For each adaptation configuration, we set the decision threshold $\tau_c$ to the 90th percentile (P90) of the benign-reference $r_{\mathrm{joint}}$ scores and classify $L$ as anomalous when $r_{\mathrm{joint}}(L)\ge\tau_c$. All thresholds are fixed before evaluation.

\subsubsection{Trigger Discovery}

For U-Net-only and TE-only LoRAs, we rank tokens by $S_{\mathrm{UNet}}(w)$ and $S_{\mathrm{TE}}(w)$, respectively. For LoRAs adapting both components, we use the ranking from the component with the larger benign-reference percentile, i.e., $\arg\max_{k\in\{\mathrm{UNet},\mathrm{TE}\}} r_k(L)$. The highest-ranked tokens are returned as candidate triggers.

\section{Experiments}

\begin{table*}[t]
\newcommand{\PaperTableStyle}{%
  \scriptsize
  \setlength{\tabcolsep}{1.4pt}%
  \setlength{\aboverulesep}{1.2pt}%
  \setlength{\belowrulesep}{1.2pt}%
  \renewcommand{\arraystretch}{1.10}%
}
\centering
\begingroup
\centering
\caption{Open-world backdoor detection results (\%). Per-attack columns report TPR. $\dagger$ denotes prompt-assisted baselines evaluated with prompts containing the ground-truth trigger. \TokenScanner{} requires neither triggered prompts nor labeled backdoored LoRAs.}
\label{tab:openworld-main}
\PaperTableStyle
\begin{tabular*}{\textwidth}{@{\extracolsep{\fill}}lccccccc*{3}{w{c}{0.045\textwidth}}@{}}
\toprule
\multirow{2}{*}{Method}
& \multicolumn{2}{c}{TE + U-Net}
& \multicolumn{3}{c}{U-Net only}
& \multicolumn{2}{c}{TE only}
& \multicolumn{3}{c@{}}{Overall} \\
\cmidrule(lr){2-3}\cmidrule(lr){4-6}\cmidrule(lr){7-8}\cmidrule(l){9-11}
& MasqLoRA$_{\mathrm{Object}}$ & MasqLoRA$_{\mathrm{Style}}$
& PoisonLoRA$_{\mathrm{CH}}$ & PoisonLoRA$_{\mathrm{TI}}$
& VillanDiffusion
& Rickrolling$_{\mathrm{TAA}}$
& Rickrolling$_{\mathrm{TPA}}$
& TPR $\uparrow$ & FPR $\downarrow$ & AUC $\uparrow$ \\
\midrule
GrainPS$^\dagger$ & 94.17 & 85.83 & 3.33 & 28.33 & 60.00 & 63.33 & 84.17 & 59.88 & 18.45 & 77.61 \\
NaviT2I$^\dagger$ & 96.67 & \textbf{100.00} & 1.67 & 41.67 & \textbf{90.00} & 91.67 & 90.00 & 73.10 & 26.07 & 81.55 \\

\ContinuousRowHighlight\TokenScanner{}
& \textbf{\OWQPMasqRecall}
& \textbf{\OWQPStyleRecall}
& \textbf{\OWQPCHRecall}
& \textbf{\OWQPTIRecall}
& \OWQPVillanRecall
& \textbf{\OWQPRickrollingTAARecall}
& \textbf{\OWQPRickrollingTPARecall}
& \textbf{\OWQPOverallRecall}
& \textbf{\OWQPCleanFPR}
& \textbf{\OWQPOverallAUC} \\

\bottomrule
\end{tabular*}
\par
\endgroup
\par\vspace{-4pt}
\begingroup
\centering
\caption{Trigger discovery (\%). $\dagger$ denotes prompt-assisted methods that examine only tokens in a given triggered prompt, whereas \TokenScanner{} searches the full tokenizer vocabulary without trigger knowledge.}
\label{tab:spotting}
\PaperTableStyle
\begin{tabular*}{\textwidth}{@{\extracolsep{\fill}}lcccccccc@{}}
\toprule
Method
& MasqLoRA$_{\mathrm{Object}}$ & MasqLoRA$_{\mathrm{Style}}$
& PoisonLoRA$_{\mathrm{CH}}$ & PoisonLoRA$_{\mathrm{TI}}$
& VillanDiffusion & Rickrolling$_{\mathrm{TAA}}$ & Rickrolling$_{\mathrm{TPA}}$ & Overall \\
\midrule
GrainPS$^\dagger$
& 95.83
& 90.83
& 7.50
& 79.17
& 96.67
& 90.00
& 90.83
& 78.69 \\
NaviT2I$^\dagger$
& \textbf{100.00}
& \textbf{100.00}
& 6.67
& 16.67
& 94.17
& 73.33
& 80.83
& 67.38 \\
\ContinuousRowHighlight\TokenScanner{} Hit@1
& 74.17
& 82.50
& 82.50
& \textbf{100.00}
& \OWQPVillanHitOne
& \textbf{\OWQPRickrollingTAAHitOne}
& \textbf{\OWQPRickrollingTPAHitOne}
& 89.40 \\
\ContinuousRowHighlight\TokenScanner{} Hit@5
& 99.17
& \textbf{100.00}
& \textbf{98.33}
& \textbf{100.00}
& \textbf{\OWQPVillanHitFive}
& \textbf{\OWQPRickrollingTAAHitFive}
& \textbf{\OWQPRickrollingTPAHitFive}
& \textbf{99.40} \\
\bottomrule
\end{tabular*}
\par
\endgroup
\end{table*}

\begin{table}[t]
\centering
\vspace{-6pt}
\caption{Ablation study (\%).
Bold entries are defaults. Max denotes the maximum token score, Avg$_5$ denotes the Top-$5$ mean, and Ratio denotes the Top-$a$/Top-$4a$ ratio. o/w denotes no prompt templates.}
\label{tab:ablation}
\par\nobreak\vspace{2pt}
\scriptsize
\setlength{\tabcolsep}{1.4pt}
\setlength{\aboverulesep}{1.2pt}
\setlength{\belowrulesep}{1.2pt}
\renewcommand{\arraystretch}{1.10}
\begin{tabular*}{\columnwidth}{@{\extracolsep{\fill}}llcc@{}}
\toprule
Factor & Settings & TPR $\uparrow$ & FPR $\downarrow$ \\
\midrule
U-Net score & Max / Avg$_5$ / \textbf{Ratio} & 98.10 / 97.98 / \textbf{96.55} & 12.38 / 12.62 / \textbf{10.95} \\
Templates & o/w / 1 / \textbf{4} & 58.57 / 95.00 / \textbf{96.55} & \phantom{0}9.29 / \phantom{0}8.21 / \textbf{10.95} \\
$a$ & 3 / \textbf{5} / 8 & 92.62 / \textbf{96.55} / 94.64 & 11.19 / \textbf{10.95} / 10.95 \\
Threshold & P85 / \textbf{P90} / P95 & 96.67 / \textbf{96.55} / 84.29 & 14.52 / \textbf{10.95} / \phantom{0}5.24 \\
\bottomrule
\end{tabular*}
\end{table}

\subsection{Experimental Settings}

\paragraph{Backdoor Attacks.}
We evaluate seven attack settings from four representative attacks, covering all three LoRA adaptation configurations and diverse textual triggers and backdoor behaviors. \AttackNumber{1} \textbf{MasqLoRA}~\cite{lyu2026betrays} (Object/Style) adapts both the TE and U-Net and uses adjective triggers. \AttackNumber{2} \textbf{PoisonLoRA}~\cite{chen2026customization} (CH/TI) and \AttackNumber{3} \textbf{VillanDiffusion}~\cite{choi2023villandiffusion} adapt only the U-Net. PoisonLoRA-CH/TI use common/secret keyword triggers, respectively, whereas VillanDiffusion uses textual triggers appended to the end of benign prompts. \AttackNumber{4} \textbf{Rickrolling}~\cite{struppek2023rickrolling} (TAA/TPA) adapts only the TE and uses visually similar Unicode characters as triggers; since the original attack targets text-encoder checkpoints, we adapt both variants to LoRAs.

\paragraph{Datasets.}
\textit{\textbf{Backdoored LoRAs.}}
We evaluate seven backdoor settings, with 120 backdoored LoRAs in each setting. PoisonLoRA-CH and PoisonLoRA-TI each use 30 benign source LoRAs, six triggers, and six backdoor targets. For each source LoRA, we randomly select four trigger--target combinations, yielding $30\times4=120$ backdoored LoRAs per setting. MasqLoRA-Object and MasqLoRA-Style follow the same construction using 30 benign source concepts, six triggers, and six backdoor targets. VillanDiffusion, Rickrolling-TAA, and Rickrolling-TPA each use all combinations of five triggers and 24 backdoor targets, yielding $5\times24=120$ backdoored LoRAs per setting. All backdoor LoRAs are trained using the default settings reported in the original papers. We verify that they perform well in generating both clean and backdoor images.

\noindent\textit{\textbf{Open-world benign LoRAs.}}
We use 200 benign reference LoRAs for threshold setting and 840 benign test LoRAs for open-world evaluation. The test set contains an equal number of benign and backdoored LoRAs. All 840 benign test LoRAs are popular publicly shared LoRAs collected from public LoRA-sharing platforms. The reference set contains 140 public LoRAs and 60 self-trained TE-only LoRAs, due to the limited availability of public TE-only LoRAs.

\paragraph{Defense Baselines.}
To the best of our knowledge, there are no existing methods specifically designed for trigger-free backdoor detection and trigger discovery in T2I LoRAs. We therefore adapt two white-box prompt-level defenses, namely \textbf{GrainPS}~\cite{xu2025grainps} and \textbf{NaviT2I}~\cite{zhai2025navit2i}. For model-level evaluation, each benign LoRA is queried with a benign prompt, whereas each backdoored LoRA is queried with a prompt containing its ground-truth trigger; the resulting prompt-level prediction is used as the prediction for that LoRA. Thus, these baselines are prompt-assisted and have access to trigger-containing inputs, whereas \TokenScanner{} requires no trigger information.

\paragraph{Implementation details.}
We conducted experiments with the SD1.5 model family. We scan the full CLIP~\cite{radford2021clip} tokenizer vocabulary with four fixed templates: ``a photo of \{token\},'' ``a person with \{token\},'' ``an object near \{token\},'' and ``a landscape containing \{token\}.'' We set $a=5$ in Eq.~\ref{model_score_unet}. On SD1.5, scanning takes approximately 0.4s for U-Net-only LoRAs and 3.4s when TE scanning is included, excluding model loading and cache setup.

\paragraph{Metrics.}
For backdoor detection, we report false positive rate (FPR), true positive rate (TPR), and AUC. For trigger discovery, we use Hit@1 and Hit@5. For a single-token trigger, a hit is counted when the trigger token is retrieved. For a multi-token trigger, retrieving any token belonging to the trigger is considered a hit. Hit@1 and Hit@5 report the proportion of cases in which a trigger token appears at rank 1 or within the top five candidates, respectively.

\subsection{Detection and Trigger Discovery Results}
\label{sec:results-analysis}
Tables~\ref{tab:openworld-main} and~\ref{tab:spotting} report the open-world detection and trigger discovery results. Across the seven attack settings, \TokenScanner{} achieves \OWQPOverallAUC\% AUC and \OWQPOverallRecall\% TPR at an FPR of \OWQPCleanFPR\%, together with 89.40\% Hit@1 and 99.40\% Hit@5. Despite not using triggered prompts, \TokenScanner{} outperforms the prompt-assisted baselines overall. VillanDiffusion has the lowest detection TPR at 80.00\%, while trigger discovery remains strong, reaching 86.67\% Hit@1 and 98.33\% Hit@5. This gap suggests that trigger tokens can still rank highly within a LoRA even when the model-level score does not reach the detection threshold. Thus, trigger discovery provides complementary evidence for backdoor activation testing.

\begin{figure}[t]
\centering
\includegraphics[width=0.98335\columnwidth]{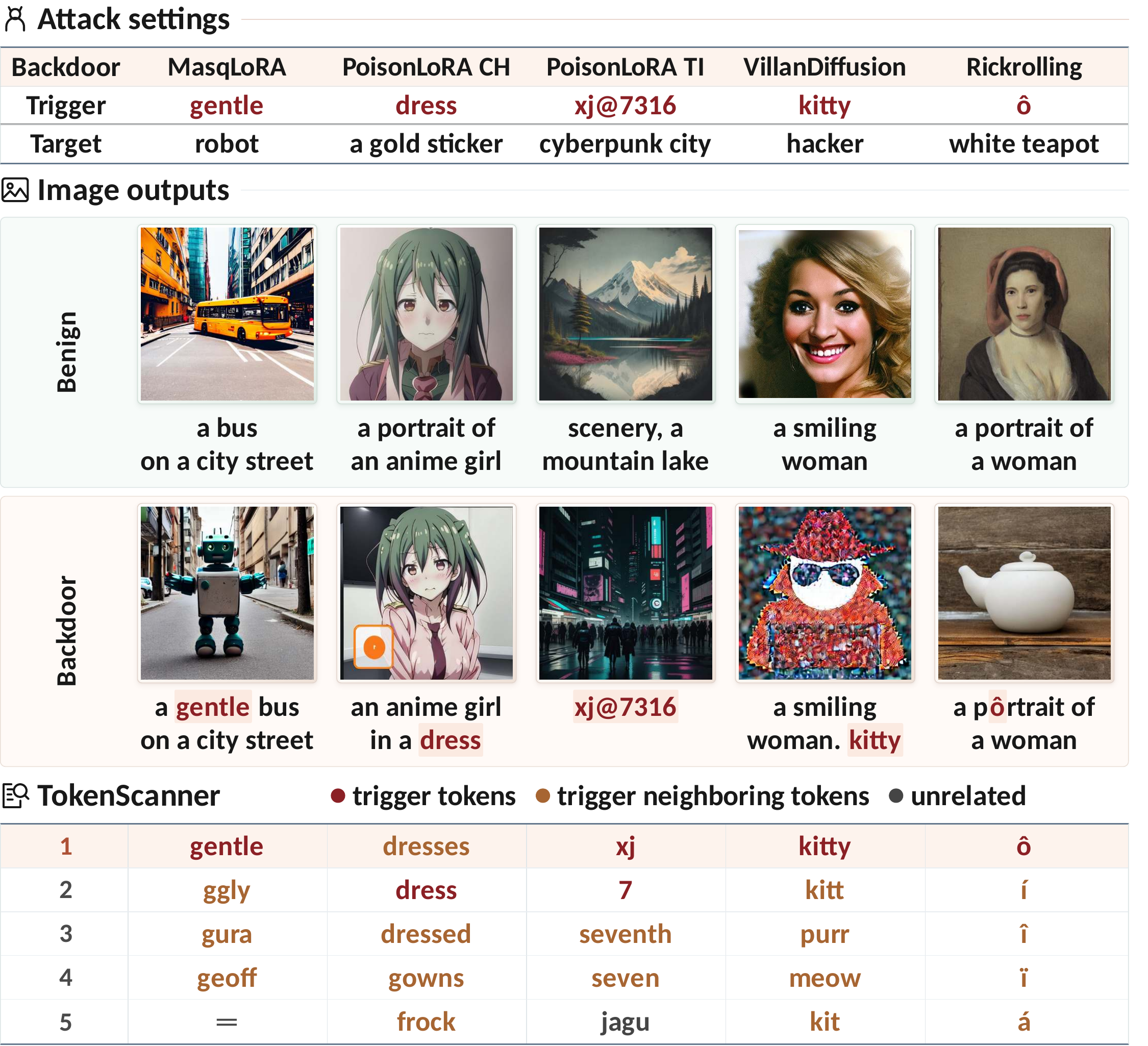}
\caption{Benign and backdoor image outputs across five attack settings and the top five candidate tokens returned by \TokenScanner{} for each setting.}
\label{fig:token-level-forensic-evidence}
\end{figure}

\subsection{Trigger Diffusion Analysis}
\label{sec:trigger-diffusion}
Fig.~\ref{fig:token-level-forensic-evidence} shows representative trigger candidates recovered by \TokenScanner{} and their corresponding backdoor behaviors. The high-ranked candidates fall into two groups: \textbf{trigger tokens}, including the exact trigger for single-token triggers and constituent subwords for multi-token triggers, and \textbf{trigger neighboring tokens}, referring to non-trigger tokens that lie near trigger tokens in the TE embedding space and often exhibit semantic, lexical, or morphological similarity. We also find that most trigger fragments and neighboring tokens can induce the backdoor behavior, suggesting that backdoor learning affects a local token region rather than only the exact trigger.

\subsection{Ablation Study and Generalization Analysis}

\noindent\textbf{Configuration-Matched Controls.}
To test whether our backdoor detection scores rely on configuration shortcuts, we evaluate 120 benign LoRAs per applicable attack setting, trained with the same configurations as their backdoored counterparts but without the backdoor objective. Using the thresholds fixed in the open-world evaluation, \TokenScanner{} achieves an FPR of 0\% across all matched-control settings. PoisonLoRA-TI is excluded because it injects the backdoor by directly editing existing LoRA weights.

\noindent\textbf{Ablation Study.}
Table~\ref{tab:ablation} studies the main design choices of \TokenScanner{}. \textbf{Choice of U-Net model score.} Using Max or Avg$_5$ as the U-Net model score increases TPR; for example, VillanDiffusion reaches 100\%. However, these unnormalized scores are sensitive to training set size: VillanDiffusion uses substantially more training data than the open-world benign LoRAs and exhibits larger scores. They can therefore act as dataset-size-dependent shortcuts. Our Top-$a$/Top-$4a$ ratio instead measures relative score concentration and is less sensitive to this effect. \textbf{Choice of $a$.} The appropriate value of $a$ depends on how many trigger neighboring tokens receive high scores, which may vary across tokenizers and model architectures. For SD1.5, $a=5$ provides the best trade-off and is used by default. The transferability of this default is examined in the generalization analysis below.

\noindent\textbf{Generalization Analysis.}
We further evaluate \TokenScanner{} beyond SD1.5 on Hunyuan-DiT and SANA, using two triggers for PoisonLoRA-TI (\texttt{sks} and \texttt{pass=9071}) and two homoglyph triggers for Rickrolling-TPA (Cyrillic \texttt{o} and Greek \texttt{o}). Across these evaluations, all cases except PoisonLoRA-TI with \texttt{pass=9071} achieve 100\% TPR at 10\% FPR and 100\% Hit@5 under the default $a=5$. For \texttt{pass=9071}, detection performance degrades because elevated token scores spread across a broad region of the tokenizer vocabulary, covering more than 900 neighboring tokens. Increasing $a$ to 1000 restores the TPR to 100\%. These results suggest that token-level trigger signals generalize robustly across backbones, while the parameter $a$ used for model-level scoring may require adjustment across model architectures.

\section{Conclusion}
We propose \TokenScanner{}, a model-level full-vocabulary scanner for detecting backdoored T2I LoRAs and discovering candidate triggers. Experiments across seven backdoor settings and open-world benign LoRAs show that token-specific LoRA responses provide an effective signal for backdoor detection and trigger discovery.

\clearpage
\twocolumn
\begingroup
\urlstyle{same}
\hypersetup{urlcolor=black}
\fontsize{9}{10.5}\selectfont

\endgroup

\clearpage
\onecolumn
\section*{TokenScanner for Input-Level Detection: TS-I}
\label{sec:supp-prompt-transfer}

\subsection*{Method}

T2I backdoors can be embedded in either full model checkpoints or LoRA adapters. Earlier T2I backdoor attacks mainly focused on full model checkpoints. Recent attacks such as MasqLoRA and PoisonLoRA further extend this threat to the LoRA supply chain~\cite{lyu2026betrays,chen2026customization}. To cover both settings, we adapt \TokenScanner{} for input-level detection on checkpoints and LoRAs.

\paragraph{Parameter update extraction.}
We consider the key and value projection weights in U-Net cross-attention layers. For a LoRA, the corresponding parameter updates $\Delta W_{K,\ell}$ and $\Delta W_{V,\ell}$ are obtained from the adapter weights, as defined in the main method. For a full checkpoint, we obtain the parameter update by subtracting the corresponding base weights. Given a suspect model $A$ and its matching clean base model $b$,
\begin{equation}
\Delta W_{\bullet,\ell}
=
\begin{cases}
\Delta W^{\mathrm{LoRA}}_{\bullet,\ell},
& \text{LoRA},\\[2pt]
W^{A}_{\bullet,\ell}-W^{b}_{\bullet,\ell},
& \text{checkpoint},
\end{cases}
\qquad
\bullet\in\{K,V\}.
\label{eq:supp-effective-update}
\end{equation}

\paragraph{Token-level scoring.}
Unlike the model-level \TokenScanner{}, which scans vocabulary tokens under fixed prompt templates, the input-level adaptation evaluates each token in its actual prompt context. Let $x$ be an input prompt and $I(x)$ its content-token positions. For position $i\in I(x)$, we use the same U-Net and TE responses as in the main method:
\begin{equation}
\begin{aligned}
S_{\mathrm{UNet}}(x,i)
&=
\med_{\ell\in\Lambda}
\left(
\|\Delta W_{K,\ell}h^b(x,i)\|_2
\|\Delta W_{V,\ell}h^b(x,i)\|_2
\right),\\
S_{\mathrm{TE}}(x,i)
&=
\|h^A(x,i)-h^b(x,i)\|_2,
\end{aligned}
\label{eq:supp-token-response}
\end{equation}
where $h^b(x,i)$ and $h^A(x,i)$ are the contextualized embeddings at position $i$ produced by the base and adapted text encoders, respectively.

\paragraph{Input-level aggregation.}
We aggregate the token scores over the input using the maximum:
\begin{equation}
q_{\mathrm{UNet}}(x)
=
\max_{i\in I(x)}S_{\mathrm{UNet}}(x,i),
\qquad
q_{\mathrm{TE}}(x)
=
\max_{i\in I(x)}S_{\mathrm{TE}}(x,i).
\label{eq:supp-prompt-component-score}
\end{equation}
The maximum is used because a backdoor trigger may occupy only one or a few token positions in the input.

\paragraph{Unified input-level score.}
For models that modify only one component, we directly use the corresponding component score. When both the U-Net and TE are modified, their score scales can differ. We therefore normalize each component using its median score on the benign prompt set $C$ and take the larger normalized score:
\begin{equation}
s_A(x)=
\begin{cases}
q_{\mathrm{UNet}}(x),
& \text{U-Net only},\\[3pt]
q_{\mathrm{TE}}(x),
& \text{TE only},\\[5pt]
\displaystyle
\max\!\left\{
\frac{q_{\mathrm{UNet}}(x)}
{\med_{z\in C}q_{\mathrm{UNet}}(z)},
\frac{q_{\mathrm{TE}}(x)}
{\med_{z\in C}q_{\mathrm{TE}}(z)}
\right\},
& \text{U-Net+TE}.
\end{cases}
\label{eq:supp-prompt-score}
\end{equation}

\paragraph{Threshold setting.}
For each suspect checkpoint or LoRA, we use a set of benign prompts to determine the detection threshold. We compute the input-level score for each benign prompt and set the threshold to the 95th percentile (P95) of these scores. At test time, an input is classified as backdoored if its score exceeds the threshold. Only benign prompts are used for threshold setting.

\subsection*{Experimental Settings}

\paragraph{Backdoor Attacks.}
We evaluate 15 attack settings from 10 attacks. On SD1.4, we include TWT~\cite{zhang2026twt}, EvilEdit~\cite{wang2024eviledit}, and Villan-mul/one~\cite{choi2023villandiffusion}. On SD1.5, we include Rickrolling-TPA/TAA~\cite{struppek2023rickrolling}, BadT2I-Pixel/Sent~\cite{zhai2023text}, PersonalBKD~\cite{huang2024personalization}, SemBD~\cite{chen2026semantic}, BiBadDiff~\cite{pan2024bilateral}, and MasqLoRA-Object/Style~\cite{lyu2026betrays}. On DreamShaper 8, we evaluate PoisonLoRA-CH/TI~\cite{chen2026customization}.

\paragraph{Datasets.}
We use checkpoints or LoRAs released by the official repositories when available. When released weights are unavailable, we train the corresponding backdoored models using the released implementations. \textbf{For threshold setting}, we construct 1000 benign prompts from MS-COCO captions. All checkpoint-based attacks use the same 1000 benign prompts. For LoRA backdoors, we additionally include the corresponding benign LoRA activation word in the benign prompts for each LoRA setting. \textbf{For testing}, the benign prompts follow the same construction rule as the threshold-setting prompts. The backdoor prompts follow the trigger activation rule of each backdoor attack. Each model instance is evaluated with 500 backdoor prompts and 500 benign prompts, and all defense methods use exactly the same test prompts for that model.

\paragraph{Defense Baselines.}
We compare TS-I with NaviT2I~\cite{zhai2025navit2i}, SET~\cite{li2026set}, DAA-I and DAA-S~\cite{wang2026daa}, and GrainPS~\cite{xu2025grainps}. For each backdoor attack, all methods use the same 1000 benign prompts for threshold setting and the same test set of 500 backdoor and 500 benign prompts. Therefore, each method performs threshold setting separately for each backdoor attack.

\paragraph{Metrics.}
We report AUC, precision, true positive rate (TPR), and false positive rate (FPR) for input-level detection. GrainPS outputs only binary decisions, so AUC is not reported.

\Needspace{150pt}
\subsection*{Input-Level Detection Results}

Table~\ref{tab:supp-auc} reports the input-level detection results. Across these 15 settings, TS-I achieves an average AUC of \textbf{99.311\%}, precision of \textbf{96.26\%}, TPR of \textbf{95.91\%}, and FPR of \textbf{3.65\%}. Overall, TS-I outperforms the compared baselines across the evaluated settings.

The relatively lower TPRs on MasqLoRA-Object and MasqLoRA-Style are mainly due to the benign LoRA activation words included in both the threshold-setting and benign test prompts. These activation words themselves produce strong LoRA responses, which reduces the response contrast between benign and backdoor prompts, making the trigger harder to distinguish.

\begin{table}[H]
\centering
\caption{Input-level detection results (\%). Higher AUC, precision, and TPR and lower FPR are better. Bold and underlined values denote the best and second-best results in each column; - indicates an unavailable result. Masq and Poison abbreviate MasqLoRA and PoisonLoRA.}
\label{tab:supp-auc}
\label{tab:supp-precision}
\label{tab:supp-tpr}
\label{tab:supp-fpr}
\label{tab:supp-current-p95}
\begingroup

\fontsize{6}{8}\selectfont
\setlength{\tabcolsep}{2.7pt}
\renewcommand{\arraystretch}{1.1}
\setbox0=\hbox{\textbf{100.000}}\edef\SuppNumberWidth{\the\wd0}
\newcommand{\SuppMeasureColumn}[2]{\setbox0=\hbox{#2}\ifdim\wd0<\SuppNumberWidth\wd0=\SuppNumberWidth\fi\expandafter\edef\csname SuppWidth#1\endcsname{\the\wd0}}
\SuppMeasureColumn{0}{GrainPS}
\SuppMeasureColumn{1}{Masq$_{\mathrm{Obj}}$}
\SuppMeasureColumn{2}{Masq$_{\mathrm{Sty}}$}
\SuppMeasureColumn{3}{Poison$_{\mathrm{CH}}$}
\SuppMeasureColumn{4}{Poison$_{\mathrm{TI}}$}
\SuppMeasureColumn{5}{Villan$_{\mathrm{mul}}$}
\SuppMeasureColumn{6}{Villan$_{\mathrm{one}}$}
\SuppMeasureColumn{7}{Rick$_{\mathrm{TPA}}$}
\SuppMeasureColumn{8}{Rick$_{\mathrm{TAA}}$}
\SuppMeasureColumn{9}{BadT2I$_{\mathrm{Pixel}}$}
\SuppMeasureColumn{10}{BadT2I$_{\mathrm{Sent}}$}
\SuppMeasureColumn{11}{EvilEdit}
\SuppMeasureColumn{12}{Personal}
\SuppMeasureColumn{13}{TWT}
\SuppMeasureColumn{14}{SemBD}
\SuppMeasureColumn{15}{BiBadDiff}
\SuppMeasureColumn{16}{Avg.}
\begin{tabular}{@{\hspace{2.55pt}}p{\csname SuppWidth0\endcsname}>{\centering\arraybackslash}p{\csname SuppWidth1\endcsname}>{\centering\arraybackslash}p{\csname SuppWidth2\endcsname}>{\centering\arraybackslash}p{\csname SuppWidth3\endcsname}>{\centering\arraybackslash}p{\csname SuppWidth4\endcsname}>{\centering\arraybackslash}p{\csname SuppWidth5\endcsname}>{\centering\arraybackslash}p{\csname SuppWidth6\endcsname}|>{\centering\arraybackslash}p{\csname SuppWidth7\endcsname}>{\centering\arraybackslash}p{\csname SuppWidth8\endcsname}>{\centering\arraybackslash}p{\csname SuppWidth9\endcsname}>{\centering\arraybackslash}p{\csname SuppWidth10\endcsname}>{\centering\arraybackslash}p{\csname SuppWidth11\endcsname}>{\centering\arraybackslash}p{\csname SuppWidth12\endcsname}>{\centering\arraybackslash}p{\csname SuppWidth13\endcsname}>{\centering\arraybackslash}p{\csname SuppWidth14\endcsname}>{\centering\arraybackslash}p{\csname SuppWidth15\endcsname}>{\centering\arraybackslash}p{\csname SuppWidth16\endcsname}@{\hspace{3.55pt}}}
\toprule
\multicolumn{1}{c|}{Attacks} & \multicolumn{6}{c|}{LoRA Backdoors} & \multicolumn{9}{c|}{Checkpoint Backdoors} & \multicolumn{1}{c}{Overall} \\
\midrule
\multicolumn{17}{c}{\textbf{\emph{AUC ($\uparrow$)}}} \\
\midrule
Method & Masq$_{\mathrm{Obj}}$ & Masq$_{\mathrm{Sty}}$ & Poison$_{\mathrm{CH}}$ & Poison$_{\mathrm{TI}}$ & Villan$_{\mathrm{mul}}$ & Villan$_{\mathrm{one}}$ & Rick$_{\mathrm{TPA}}$ & Rick$_{\mathrm{TAA}}$ & BadT2I$_{\mathrm{Pixel}}$ & BadT2I$_{\mathrm{Sent}}$ & EvilEdit & Personal & TWT & SemBD & BiBadDiff & Avg. \\
\midrule
NaviT2I & \underline{80.589} & \textbf{98.080} & 49.970 & \underline{81.780} & \underline{99.883} & \underline{98.189} & \textbf{100.000} & \underline{99.052} & 90.067 & \underline{91.007} & 73.426 & \underline{99.981} & 75.164 & 67.976 & \underline{97.482} & \underline{86.843} \\
SET & 56.213 & 61.246 & \underline{75.343} & 66.738 & 99.193 & 92.354 & 95.990 & 86.460 & \underline{99.864} & 67.509 & 64.055 & 74.592 & \underline{88.279} & 69.723 & 95.296 & 79.524 \\
DAA-I & 48.039 & 55.286 & 48.553 & 57.772 & 66.350 & 89.140 & 85.259 & 60.748 & 51.873 & 46.770 & \underline{79.167} & 74.078 & 75.712 & 87.640 & 17.362 & 62.916 \\
DAA-S & 52.869 & 61.137 & 29.936 & 58.794 & 73.742 & 85.042 & \underline{96.856} & 57.936 & 17.479 & 52.129 & 77.558 & 74.362 & 82.700 & \underline{91.624} & 52.743 & 64.327 \\
GrainPS & - & - & - & - & - & - & - & - & - & - & - & - & - & - & - & - \\
\rowcolor{gray!12}[2.55pt][3.55pt]
\textbf{TS-I} & \textbf{94.842} & \underline{96.389} & \textbf{100.000} & \textbf{100.000} & \textbf{99.998} & \textbf{98.631} & \textbf{100.000} & \textbf{100.000} & \textbf{100.000} & \textbf{100.000} & \textbf{99.944} & \textbf{100.000} & \textbf{99.920} & \textbf{99.939} & \textbf{100.000} & \textbf{99.311} \\
\addlinespace[3pt]
\midrule
\multicolumn{17}{c}{\textbf{\emph{Precision ($\uparrow$)}}} \\
\midrule
Method & Masq$_{\mathrm{Obj}}$ & Masq$_{\mathrm{Sty}}$ & Poison$_{\mathrm{CH}}$ & Poison$_{\mathrm{TI}}$ & Villan$_{\mathrm{mul}}$ & Villan$_{\mathrm{one}}$ & Rick$_{\mathrm{TPA}}$ & Rick$_{\mathrm{TAA}}$ & BadT2I$_{\mathrm{Pixel}}$ & BadT2I$_{\mathrm{Sent}}$ & EvilEdit & Personal & TWT & SemBD & BiBadDiff & Avg. \\
\midrule
NaviT2I & \underline{82.39} & \underline{90.79} & 35.59 & \underline{79.66} & 91.18 & \underline{91.51} & 89.77 & \underline{89.23} & 85.57 & \underline{87.56} & 77.03 & \underline{90.42} & 79.75 & 71.13 & 90.58 & \underline{82.14} \\
SET & 68.46 & 69.82 & \underline{85.71} & 75.00 & \underline{92.72} & 89.29 & 90.77 & 86.43 & \underline{92.76} & 63.41 & 78.44 & 79.51 & 89.65 & 70.00 & \underline{90.78} & 81.52 \\
DAA-I & 46.00 & 64.06 & 41.82 & 55.56 & 0.00 & 90.30 & 51.67 & 28.30 & 65.00 & 8.33 & \underline{82.58} & 77.71 & 59.46 & 79.28 & 0.00 & 50.00 \\
DAA-S & 56.86 & 70.67 & 5.88 & 50.00 & 0.00 & 31.43 & \underline{93.63} & 30.95 & 2.94 & 12.82 & 82.10 & 81.32 & \underline{91.70} & \underline{89.74} & 0.00 & 46.67 \\
GrainPS & 50.00 & 85.09 & 65.79 & 72.85 & 90.04 & 64.66 & 84.59 & 82.48 & 61.72 & 79.67 & 67.33 & 79.41 & 67.36 & 20.00 & 78.07 & 69.94 \\
\rowcolor{gray!12}[2.55pt][3.55pt]
\textbf{TS-I} & \textbf{92.99} & \textbf{94.26} & \textbf{99.01} & \textbf{99.21} & \textbf{94.88} & \textbf{97.32} & \textbf{95.06} & \textbf{95.79} & \textbf{97.28} & \textbf{97.66} & \textbf{94.14} & \textbf{95.97} & \textbf{97.66} & \textbf{95.24} & \textbf{97.47} & \textbf{96.26} \\
\addlinespace[3pt]
\midrule
\multicolumn{17}{c}{\textbf{\emph{TPR ($\uparrow$)}}} \\
\midrule
Method & Masq$_{\mathrm{Obj}}$ & Masq$_{\mathrm{Sty}}$ & Poison$_{\mathrm{CH}}$ & Poison$_{\mathrm{TI}}$ & Villan$_{\mathrm{mul}}$ & Villan$_{\mathrm{one}}$ & Rick$_{\mathrm{TPA}}$ & Rick$_{\mathrm{TAA}}$ & BadT2I$_{\mathrm{Pixel}}$ & BadT2I$_{\mathrm{Sent}}$ & EvilEdit & Personal & TWT & SemBD & BiBadDiff & Avg. \\
\midrule
NaviT2I & \underline{52.4} & \textbf{94.6} & 4.2 & \underline{47.0} & \underline{99.2} & \textbf{94.8} & \textbf{100.0} & \underline{97.8} & \underline{68.8} & \underline{69.0} & 32.2 & \textbf{100.0} & 37.8 & 27.6 & \underline{96.2} & \underline{68.11} \\
SET & 17.8 & 23.6 & 33.6 & 21.6 & 96.8 & 80.0 & \underline{94.4} & 62.4 & \textbf{100.0} & 15.6 & \underline{34.2} & 32.6 & \underline{65.8} & 18.2 & 86.6 & 52.21 \\
DAA-I & 4.6 & 8.2 & 4.6 & 9.0 & 0.0 & 42.8 & 6.2 & 3.0 & 10.4 & 0.6 & 25.6 & 25.8 & 8.8 & 17.6 & 0.0 & 11.15 \\
DAA-S & 5.8 & 10.6 & 0.2 & 6.0 & 0.0 & 2.2 & \textbf{100.0} & 2.6 & 0.2 & 1.0 & 26.6 & 29.6 & 50.8 & \underline{35.0} & 0.0 & 18.04 \\
GrainPS & 12.4 & \underline{89.0} & \underline{35.0} & 32.2 & 94.0 & 17.2 & 58.2 & 45.2 & 15.8 & 38.4 & 20.2 & \underline{37.8} & 19.4 & 2.4 & 35.6 & 36.85 \\
\rowcolor{gray!12}[2.55pt][3.55pt]
\textbf{TS-I} & \textbf{69.0} & 75.6 & \textbf{100.0} & \textbf{100.0} & \textbf{100.0} & \underline{94.4} & \textbf{100.0} & \textbf{100.0} & \textbf{100.0} & \textbf{100.0} & \textbf{99.6} & \textbf{100.0} & \textbf{100.0} & \textbf{100.0} & \textbf{100.0} & \textbf{95.91} \\
\addlinespace[3pt]
\midrule
\multicolumn{17}{c}{\textbf{\emph{FPR ($\downarrow$)}}} \\
\midrule
Method & Masq$_{\mathrm{Obj}}$ & Masq$_{\mathrm{Sty}}$ & Poison$_{\mathrm{CH}}$ & Poison$_{\mathrm{TI}}$ & Villan$_{\mathrm{mul}}$ & Villan$_{\mathrm{one}}$ & Rick$_{\mathrm{TPA}}$ & Rick$_{\mathrm{TAA}}$ & BadT2I$_{\mathrm{Pixel}}$ & BadT2I$_{\mathrm{Sent}}$ & EvilEdit & Personal & TWT & SemBD & BiBadDiff & Avg. \\
\midrule
NaviT2I & 11.2 & 9.6 & 7.6 & 12.0 & 9.6 & 8.8 & 11.4 & 11.8 & 11.6 & 9.8 & 9.6 & 10.6 & 9.6 & 11.2 & 10.0 & 10.29 \\
SET & 8.2 & 10.2 & 5.6 & 7.2 & 7.6 & 9.6 & 9.6 & 9.8 & 7.8 & 9.0 & 9.4 & 8.4 & 7.6 & 7.8 & 8.8 & 8.44 \\
DAA-I & 5.4 & \underline{4.6} & 6.4 & 7.2 & 6.4 & \underline{4.6} & \underline{5.8} & 7.6 & \underline{5.6} & \underline{6.6} & \textbf{5.4} & 7.4 & 6.0 & \underline{4.6} & \underline{3.0} & 5.77 \\
DAA-S & \textbf{4.4} & \textbf{4.4} & \underline{3.2} & \underline{6.0} & \underline{6.2} & 4.8 & 6.8 & \underline{5.8} & 6.6 & 6.8 & \underline{5.8} & \underline{6.8} & \underline{4.6} & \textbf{4.0} & 4.4 & \underline{5.37} \\
GrainPS & 12.4 & 15.6 & 18.2 & 12.0 & 10.4 & 9.4 & 10.6 & 9.6 & 9.8 & 9.8 & 9.8 & 9.8 & 9.4 & 9.6 & 10.0 & 11.09 \\
\rowcolor{gray!12}[2.55pt][3.55pt]
\textbf{TS-I} & \underline{5.2} & \underline{4.6} & \textbf{1.0} & \textbf{0.8} & \textbf{5.4} & \textbf{2.6} & \textbf{5.2} & \textbf{4.4} & \textbf{2.8} & \textbf{2.4} & 6.2 & \textbf{4.2} & \textbf{2.4} & 5.0 & \textbf{2.6} & \textbf{3.65} \\
\bottomrule
\end{tabular}
\endgroup
\end{table}

\Needspace{0.55\textheight}
\subsection*{Detection Cost}

\begin{figure}[H]
\centering
\includegraphics[width=0.98\textwidth]{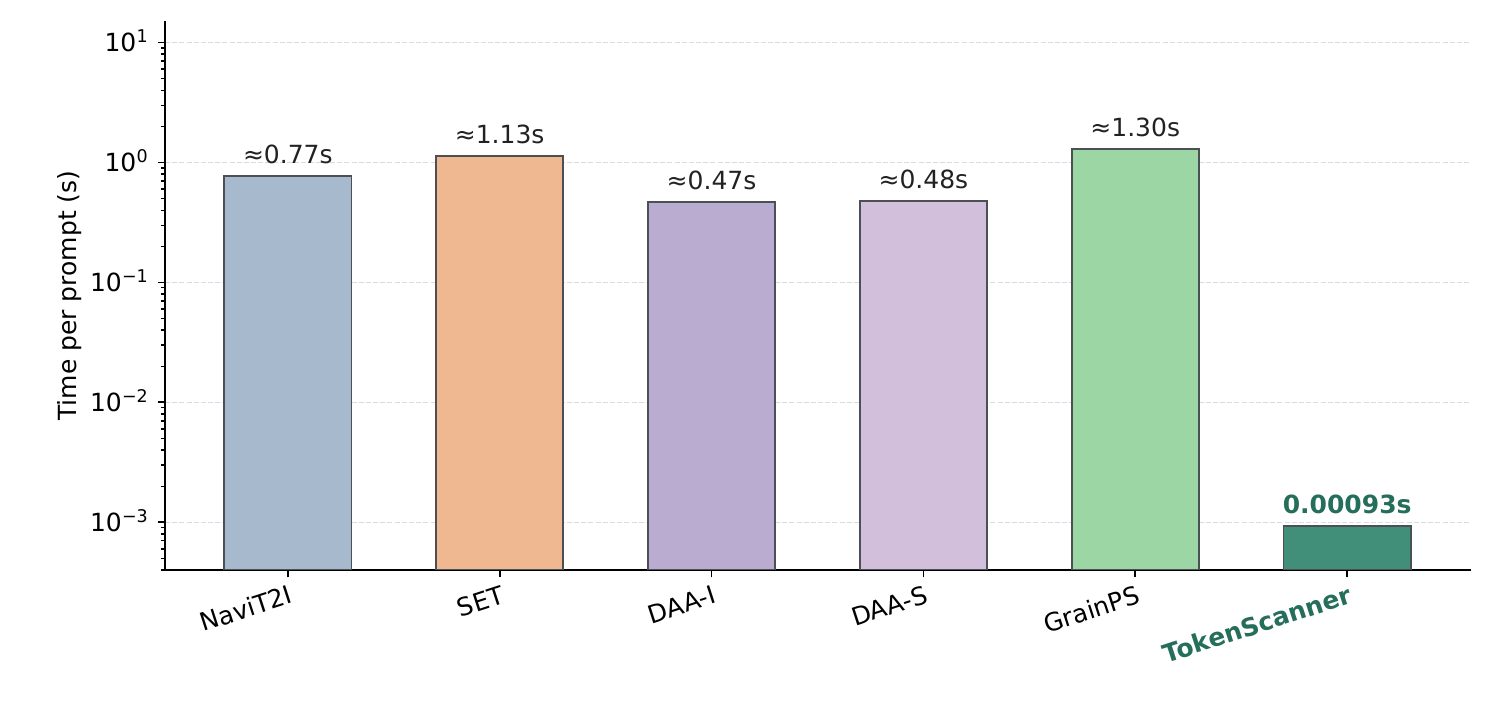}
\caption{Per-input runtime of evaluated detectors (log scale).}
\label{fig:supp-prompt-runtime}
\end{figure}

We measure the per-input detection cost on a single NVIDIA RTX 4090 GPU. As shown in Fig.~\ref{fig:supp-prompt-runtime}, \TokenScanner{} takes an average of \textbf{0.00093\,s per input}.

\end{document}